\documentclass[sigconf]{acmart}
\AtBeginDocument{%
  \providecommand\BibTeX{{%
    \normalfont B\kern-0.5em{\scshape i\kern-0.25em b}\kern-0.8em\TeX}}}

\setcopyright{acmcopyright}
\copyrightyear{2021}
\acmYear{2021}

\acmConference[EdgeSys '21]{Edinburgh '21: EdgeSys}{ Apr 26, 2021}{Edinburgh, UK}

\definecolor{electricviolet}{rgb}{0.56, 0.0, 1.0}
\usepackage{enumitem}
    \setlist[enumerate]{leftmargin=*}%
    \setlist[itemize]{leftmargin=*}%

\begin{document}

\title{Disaggregated Memory at the Edge}


\author{Luis M Vaquero}
\email{luis.vaquero@bristol.ac.uk}
\affiliation{%
  \institution{University of Bristol}
  \country{United Kingdom}
}

\author{Yehia Elkhatib}
\email{y.elkhatib@lancaster.ac.uk}
\affiliation{%
  \institution{University of Glasgow}
  \country{United Kingdom}
}

\author{Felix Cuadrado}
\email{felix.cuadrado@upm.es}
\affiliation{%
  \institution{Universidad Politecnica de Madrid}
  \country{Spain}
}

\renewcommand{\shortauthors}{Vaquero et al.}

\begin{abstract}
  This paper describes how to augment techniques such as Distributed Shared Memory with recent trends on disaggregated Non Volatile Memory in the data centre so that the combination can be used in an edge environment with potentially volatile and mobile resources.  This article identifies the main advantages and challenges, and offers an architectural evolution to incorporate recent research trends into production-ready disaggregated edges. We also present two prototypes showing the feasibility of this proposal.
\end{abstract}

\ccsdesc[500]{Applied computing}
\ccsdesc[500]{Hardware~Emerging architectures}

\keywords{edge, cloud, disaggregation, NVM}


\maketitle

\section{Introduction}



\textbf{Resource disaggregation} started with storage (e.g. storage volumes are dynamically attached to virtual machines), but it has also reached memory systems~\cite{Lim2012}. Memory disaggregation makes idle memory available to other nodes; this available memory can be from  the  same  physical  node (node  level  memory  disaggregation)  or  from  remote  nodes  in  the  same cluster (cluster level memory disaggregation)~\cite{Liu2019}. New non-volatile memory (NVM) and optical communication technologies are making it possible to realise the vision of disaggregated memory in the data centre~\cite{Faraboschi2015, vaquero2020}.   



One key element in data centre disaggregation have been NVM technologies which enable lower energy consumption and the ability to preserve state independently of compute nodes. These advantages fit the idea of the Internet-of-Things (IoT) and the resource constraints of most edge devices~\cite{Georgakopoulos2016}. Our work takes memory disaggregation to a new level by expanding it to the edge (edge level disaggregation). 



Beyond the boundaries of the data centre, latency limits are one of the drivers towards smaller units of computation and disaggregated devices and volatility and reliability call for NVM as a safe harbour for data. Also, on the privacy front, with increasing awareness and regulation, the most cost-effective mechanism of all is not to store/process data centrally. The ability to dynamically and seamlessly integrate edge memory nodes to provide persistence (via NVM exposed as a service) resources to moving and potentially battery limited compute nodes could simplify handling of state and benefiting from physical locality. 


 
To illustrate the potential of disaggregated NVM at the edge, let us imagine each fixed element in a street (e.g. buildings, lamp posts, etc.) has its own addressable non-volatile memory including its GPS coordinates and a specification of its spatial position, volume, and other physical properties. This information could be used by:

\begin{itemize}
    \item A self-driven vehicle to reconstruct the scene with much less required computation/sensors, better performance under adverse weather conditions, and simpler (energy efficient) machine learning models. 
    \item An advertisement company trying to show interactive holographic ads to pedestrians walking in the area. Some of the elements of the ads to be projected could be pre-stored in specific memory locations on the edge (based on the angle of approach ). 
\end{itemize}

These two use cases highlight the need for ultra low latency (preload of some frames across local memory locations and direct memory access for the nearby projectors/cars) to reconstruct the holographic image or a map of obstacles. They also show how NVM access can help expedite local computations in the edge. 

Many businesses such as energy, airlines, and hotel markets exhibit substantial demand fluctuations and high capacity costs. Increasing utilisation via disaggregation of resources and higher reliance on edge deployment becomes a strong economic incentive~\cite{Xu2013pricing,Kilcioglu2017}. 

Here, we present our effort towards building a new breed of systems that brings the advantage of data centre memory disaggregation to the edge. \textbf{Disaggregated edges} provide consistent, low-latency, distributed access to data stored in distributed NVM by addressing it with no centralised control. The remainder of this paper is organised as follows: Section~\ref{sec:sota} presents related work that is a precursor of disaggregated edges. In Section~\ref{sec:proposal} we introduce our framework to enable dynamically expandable virtual memory space seamless integration of edge devices. Then, two prototypes are described that demonstrate the feasibility of this approach (see Section \ref{sec:proto}). Section \ref{sec:discuss} discusses the main pros and cons of our framework in the context of recent work and we summarise the main conclusions in Section~\ref{sec:conc}.



\section{Related Work}
\label{sec:sota}
Our work builds on the concept of distributed shared memory, adapting it to edge systems that are mobile and of highly volatile resources.


\subsection{Distributed Shared Memory}

Software distributed shared memory (DSM) systems provide shared memory abstractions for clusters. DSM systems have long held the promise of a distributed virtual memory space that enables several processors to share data with each other. Confined to a cluster in a data centre, traditional DSMs rely on static address spaces (e.g the classic partitioned global address space PGAS~\cite{pgas}), where each address is a tuple of a rank in the job (or global process identifier) and an address within that same process. Dynamic and expandable address spaces will be essential to enable dynamic virtual memory space integration across edge devices. 

Our work builds on the notions of request-centric DSMs and, thus, shares some of its challenges (such as consistency, coherence, latency). At the same time, adding DSM to mobile and volatile edge resources imposes a few additional challenges: 

\begin{itemize}
    \item Memory management modules (MMMs) in DSMs do not route requests and learn memory addresses dynamically.
    \item The links between different MMMs are predefined  and often hardwired and confined to nearby processors in most DSM systems.
    \item The size of the virtual memory space is not dynamic and new nodes cannot join in and out of the shared memory space as they move around.
\end{itemize}

While some works have explored DSMs built in at a web browser level~\cite{Teragni2020}, latency has been a traditional problem for DSMs to function at scale~\cite{Nelson2015}. It builds up when working at the top of the stack and byte addressable load and store operations are not possible. 

Edge scenarios beget unpredictability; mobility, geographical location and its effect on the network and reliability would advise to minimise the encapsulation and system calls required to make remote bytes available as if they were local. Hence our work explores techniques that are closer to managing specific hardware (e.g. bespoke NVMs).

\subsection{Disaggregated Memory in the Data Centre}


Memory disaggregation  detaches   physical   memory   allocated  to  virtual  servers  at   initialisation  from  the  runtime  management  of  the  memory~\cite{Liu2019}, so that the virtual servers can use idle local (from  the  same  physical  node - known as node  level  memory  disaggregation)  or  from  remote  nodes  in  the  same cluster (known as cluster level memory disaggregation)~\cite{Liu2019}. The present work takes memory disaggregation to a new level by expanding it to the edge (we refer to this as edge level disaggregation). 


Accessing NVM by several compute units requires coherence protocols. This challenge is not new since NICs and GPUs do not maintain cache coherence with CPUs. 
Many multi-core processors have already demonstrated the use of non-coherent shared memory. Data is usually accessed through some form of logical hierarchy that helps to support access rights and multi-tenancy. Thus, a software booking system, in the form of a distributed memory controller, is required (e.g. HPE's `librarian'\footnote{https://github.com/FabricAttachedMemory/}). \cite{Faraboschi2015, vaquero2020} propose data centre level disaggregation with a routable protocol to direct load and store operations to the right machine in the data centre. In contrast,  \cite{Liu2019} use non-routable remote memory access, RDMA, for memory disaggregation. We build on routable memory abstractions but take this ability beyond the boundaries of the data centre. 

There is experimental evidence backing up the feasibility of the memory disaggregated, accelerated, optical data center. These prototypes apply a few changes to the hypervisor and operating system~\cite{caldwell2017fluidmem}, or propose a hierarchical orchestration of memory resources in a cluster~\cite{xmempod}. Rather than working within the boundaries of a data centre, our work builds on the self-organising principles that govern routed networks 
to help nodes in the edge self-organise in the creation of a dynamic virtual memory space.


\subsection{Distributed OS for Disaggregated Hardware}

\cite{Moreno-Vozmediano2019} centralise application management of edge deployed containers. In their work disaggregation does not refer to individual hardware resources.

Going one step further into hardware disaggregation without central coordination, edge devices have discovery and negotiation `agents' that enable them to request distributed resources and expose them to the applications as if they were local. These agents can be part of a decomposed operating system~\cite{Shan2018} and supported by local purpose-specific accelerators. The `software-defined machine'~\cite{Truong2015} is one of the precursors of this type of system. Our work expands on these ideas to enable a software-defined dynamic virtual memory space. Unlike many recent approaches (see \cite{Zavodovski19} for example), we do not assume the edge is just a nearby cloud where resources can be assigned and tasks run. 


\section{Disaggregated Edge}
\label{sec:proposal}
We now give an overview of our framework design and its internal mechanisms.

\subsection{Guiding Principles}

Edge resources are inherently different to cloud resources in that they tend to be more volatile; the nodes they serve are often mobile and their presence in the physical neighbourhood of the edge is ephemeral. 
This paper proposes flattening the communication stack to move data between edge devices. While traditional computing moves data to the nearby compute unit, or compute to pre-existent data, our proposal is looking for a middle ground. 

In our model data is persisted in NVM and can be dynamically accessed by passing-by devices, which seamlessly and dynamically federate their virtual memory address space with that exposed by nearby NVM modules. 

NVM modules in close proximity participate in the creation of a mesh/overlay. This is a shared virtual memory space, rather than a set physical topology.

Any device can access local NVM modules as part of their memory hierarchy in a classic, static way (as NVM RAM memory). Devices can also access memory that is remote to its local machine via a set of protocols, the local operating system expanding its virtual memory hierarchy to consider remote memory modules as if they were local. Memory requests from the CPU are routed to the appropriate remote module or a gateway memory module.

Our work assumes memory requests are sent over wireless protocols such as Zigbee, although other wireless  (e.g. Z-wave) or wired communication protocols (see Gen-Z Phy\footnote{https://genzconsortium.org/specification/gen-z-physical-layer-specification-1-1/}) would also be employable.


\subsection{Design}

\begin{figure}[th]
\centering
\includegraphics[width=0.8\columnwidth]{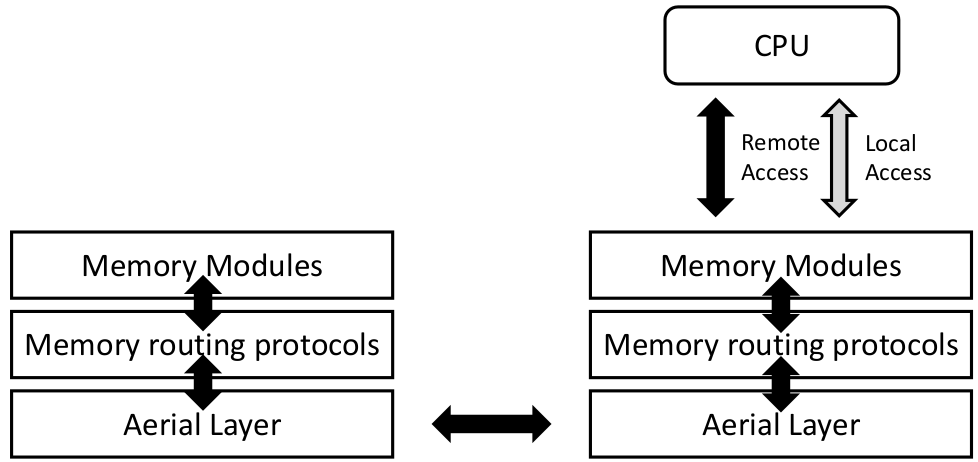}
\caption{Architectural View of the Proposed System: including hardware elements (memory modules and aerial layer/antennas and transducers), protocols (exchanges of messages for routing or for data transmission), and interactions using those protocols (arrows). }
\label{fig:arch}
\end{figure}

Figure \ref{fig:arch} describes a high level view of the architectural elements of a disaggregated edge. As can be observed, it consists of: 1) An aerial (wireless) layer in charge of encapsulating memory load/store requests and delivering them to memory modules in the range of the sender memory block. 2) A set of memory modules in charge of storing the data and converting aerial messages to electrical signals. 3) Memory routing protocols (predefined sequences of messages exchanged by memory modules) in charge of forwarding memory load/store requests through the appropriate memory module in range. Note that routing towards devices not in the immediate range of the emitter memory module is possible thanks to these set of routing protocols.

Memory routing protocols are in charge of 3 fundamental tasks:
\begin{itemize}
    \item Discovery of nearby available NVM modules that can be dynamically and seamlessly integrated into the local virtual memory so that the local operating system sees it as if it were local. These  protocols are also in charge of creating a unified virtual memory address space between all the modules participating in the mesh (see Figure~\ref{fig:mesh}). 
    \item Memory routing protocols for delivering load and store requests into a remote memory module and retrieve the data.
    \item Data coherence protocols that cope with several CPUs accessing the same remote memory (several implementations are possible: e.g. assuming immutable data and versioning for updates, distributed locks, etc.).
\end{itemize}

In our initial implementation of the aerial layer, we  encapsulated classic load/store memory primitives into the ZigBee protocol. Zigbee is defined by layer 3 and above and relies on 802.15.4 for layers 1 and 2. Zigbee allows us to create mesh topologies for devices within range, so that any node can communicate with any other node either directly or by relaying the transmission through multiple additional memory modules.

We have defined a hardware memory interface to enable memory modules to interact with each other and create a virtual address space. This interface receives aerial layer load/store requests and extracts them as a payload of an aerial protocol such as ZigBee. 

\subsection{Memory Management}

The memory management protocol is based on the Gen-Z standard, enabling memory modules to join/leave and discover nearby memory modules, "the goal of expanding a simple compute node with additional fabric-based memory, storage, networking, or accelerators"~\footnote{https://genzconsortium.org/} which itself relies on Distributed Management Task Force's Redfish~\footnote{https://www.dmtf.org/standards/redfish}. This subsection defines some distinctive elements our memory modules had to implement to perform several of the roles defined in the Redfish specification (mainly the Gen-Z Fabric Manager).

\begin{figure}[th]
\centering
\includegraphics[width=0.7\columnwidth,clip,trim=3cm 2.5cm 8cm 1cm]{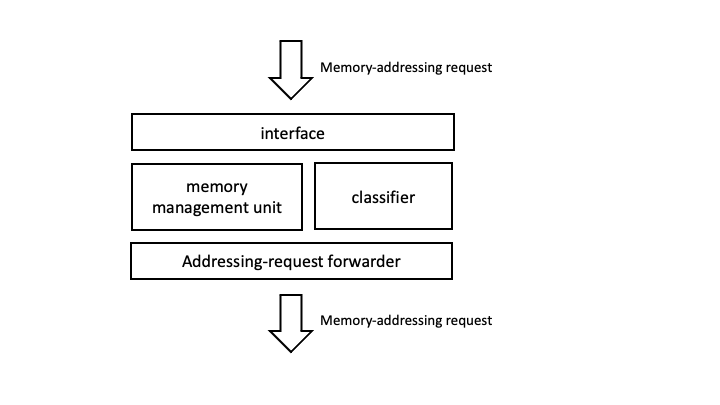}
\caption{Logical view of the hardware components of a memory router}
\label{fig:mem}
\end{figure}

Figure \ref{fig:mem} shows the main components of the memory module. Memory modules can act as destinations or relays. They consist of 1) an interface for receiving memory-addressing requests from other modules in the memory fabric or end devices, and 2) an addressing-request forwarder to control routing of memory-addressing requests. The interface extracts load/store operations from the Zigbee message and passes it on to the forwarder.

The forwarder accesses a dynamically created overlay routing table to determine a value/cost associated with inter-overlay-node paths and make routing decisions. The interface enables a number of memory modules to be connected (10 in its current implementation), effectively forming a mesh of memory modules that are directly connected to each other. 

\begin{figure}[h]
\centering
\includegraphics[width=0.2\textwidth]{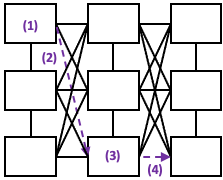}
\caption{Mesh of hardware modules~\cite{vaquero2020}}
\label{fig:mesh}
\end{figure}

The forwarder accesses a memory management unit in charge of maintaining a list of modules who are part of the same overlay (thus creating a shared virtual memory space).

Overlay routers and membership managers may exchange addressing requests over the overlay forwarding mesh itself, rather than over direct in-memory paths. This way, even if some underlying fabric paths fail these messages can still be forwarded. 
Figure~\ref{fig:mesh} shows how a load and store operation from the top left memory module \textcolor{electricviolet}{\textbf{(1)}} is forwarded to a nearby (i.e., within wireless range) module \textcolor{electricviolet}{\textbf{(2)}}, which decides the best route \textcolor{electricviolet}{\textbf{(3)}} and forwards it to \textcolor{electricviolet}{\textbf{(4)}}, its final destination. 

The memory module supports a traffic classification element that enables traffic to be routed over preferential paths (e.g. low congestion, high throughput, etc.) supporting quality of service in some memory requests.


Virtual addresses in this vast and dynamically federated memory fabric (modules can come in and out of the overlay) are unique and consist of a MAC address uniquely identifying the module and a number of bits that depends on the capacity of the modules in the mesh. This way, the MAC address can be used to route memory requests and an offset in bytes is then applied over the capacity of that memory module~\cite{vaquero2020}.

\section{Prototypes}
\label{sec:proto}
This section presents two prototypes, with two clear objectives, namely to demonstrate that it is possible to effectively (1) create a dynamically sized virtual memory federation and (2) route messages between NVM modules in mobile targets.  

\subsection{Persisting Configuration in Remote NVM}

Reducing the power consumption of a modern building requires continuous monitoring of various environmental parameters inside and outside the building. The key requirement for efficient monitoring and controlling is that all sensors and actuators are addressable over the network. 

Smart buildings include highly configurable setups for thermostats, personal voice assistants and also lighting. In commercial settings light is used to deliver marketing messages or affect shoppers behaviour. In this experiment, we wanted to ensure commercial messages can be delivered as a set of bulbs deployed in the glass window of a shop and, eventually, a skyscraper. 

We attached a set of colour-configurable bulbs to a Raspberry Pi running a modified version of Ubuntu~v16.04. 

We created a virtual address consisting of one more bit than the number of available hardware bits. When this additional virtual bit was used, a remote location was employed. In practice, we simply redirected addresses corresponding to the last addressable bit to point to the remote record in the Raspberry Pi. 

We assumed a simple protocol by which devices configure each other with a colour code to change the light of the bulb and a timestamp (the latest version of the configuration prevails). Here, we used an implementation of the NVM hardware memory module. 

\begin{figure}[ht]
\centering
\includegraphics[width=\columnwidth]{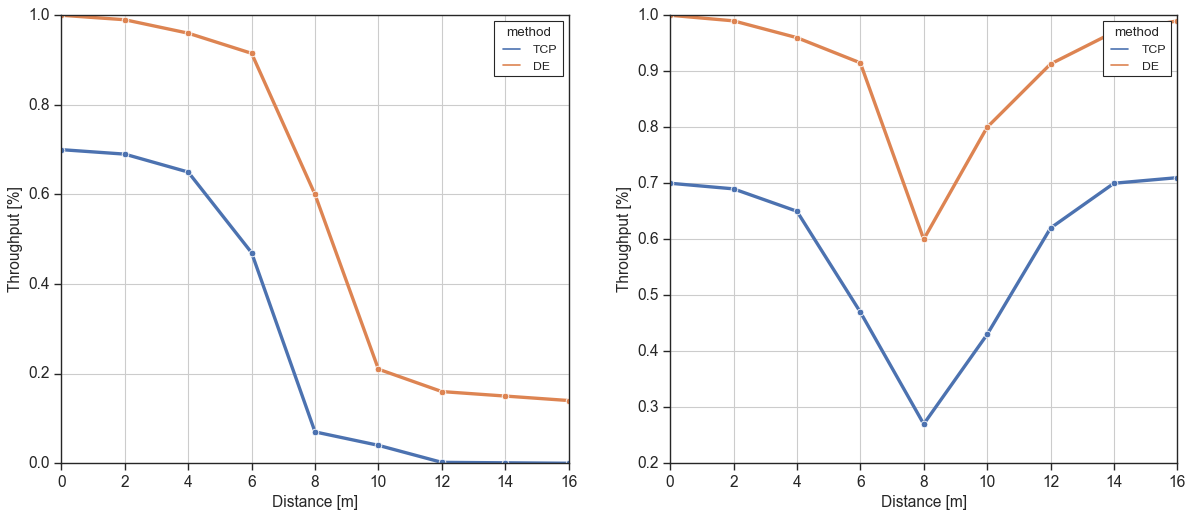}
\caption{Comparison of the throughput obtained for a single static NVM module (left) and two NVM modules (right) set 16m apart for Disaggregated Edge protocols (simplified Gen-Z over Zigbee) vs. TCP over WiFi.}
\label{fig:static}
\end{figure}

As shown in Figure ~\ref{fig:static}, a simple implementation of our protocol delivers better performance than equivalent TCP connections. The panel on the left hand side shows the throughput with a single NVM: as distance from the NVM increases, throughput decreases to a point that makes communications unfeasible. Adding a second NVM 16m away (shown on the right hand side panel) means there is a valley of performance nearly half way between the two antennas.

\subsection{Reduced Computations on Car Simulations}

We also want to investigate how a fully working prototype would behave in a more dynamic environment. 

\begin{figure}[ht]
\centering
\includegraphics[width=6cm]{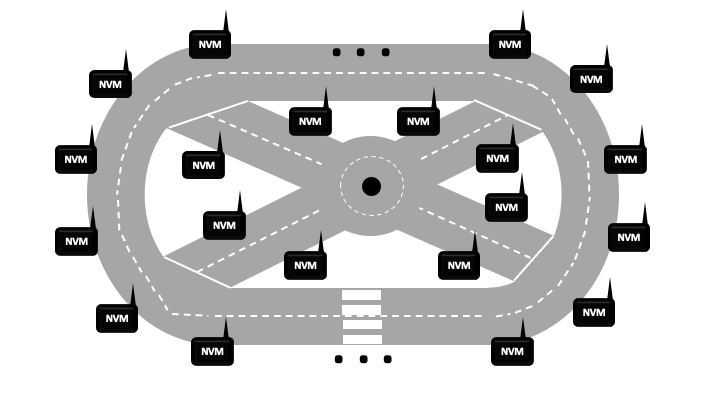}
\caption{A simple setup for simulating an urban environment with fixed elements (e.g. lamp posts) including NVM modules.}
\label{fig:setup}
\end{figure}

We aimed to simulate urban elements that would not need to be recomputed every time (e.g. buildings and other non-movable obstacles). Hence, as shown in Figure \ref{fig:setup}, we used a set of 30 Raspberry Pis with attached SD cards and a similarly modified version of Ubuntu~v16.04 (labelled as NVM in the figure). We placed them in strategic locations in a room with sticky tape defining the lane limit lines for a $\frac{1}{10}$th scaled self-driving vehicle. 

In this prototype, we also used a modified version of the scaled self-driven vehicle  by \cite{OKelly2019}, which includes cameras and a Lidar and our modified Ubuntu~v16.04 running on a Jetson TX2 device (not represented in Figure \ref{fig:setup}).

As the car moves along the track, the simulated NVM module (SD card) on the Jetson TX2 tries to sync up with nearby memory modules to become part of the memory fabric. The car tries to load information for bytes associated with a fixed standpoint tag, which then can be used to identify obstacles as the car drives along the track.

We implemented a convolutional neural network and concatenated the image input with an embedding of the obstacles declared in the NVM modules representing route obstacles. The proposed method enables the reinforcing learning algorithm to process 60\% less information on a shallower neural architecture as the algorithm can read static road properties (bends, trees, lamp posts, etc.) directly from memory. 

The simulated obstacles can be changed dynamically by simply updating the obstacle coordinates in the track side NVM.

The model is trained under the Q-learning algorithm. After continuous training for 240 minutes, the model learns the control policies to stay on track and avoid changing obstacles. 
We placed the car on the track (schematic shown in Figure \ref{fig:setup}) in order to check the behaviour of the car a dynamically federated virtually memory space. 

Latency stays under 200 ms for payloads under 200 bytes and less than 5 hops between MMs ($90_{th}$ percentile). While latency tends to be a dominant problem in the edge, high throughput is essential for self-driving cars. As the device moves, the re-connections and re-sending of data can severely affect throughput.

\begin{figure}[ht]
\centering
\includegraphics[width=\columnwidth]{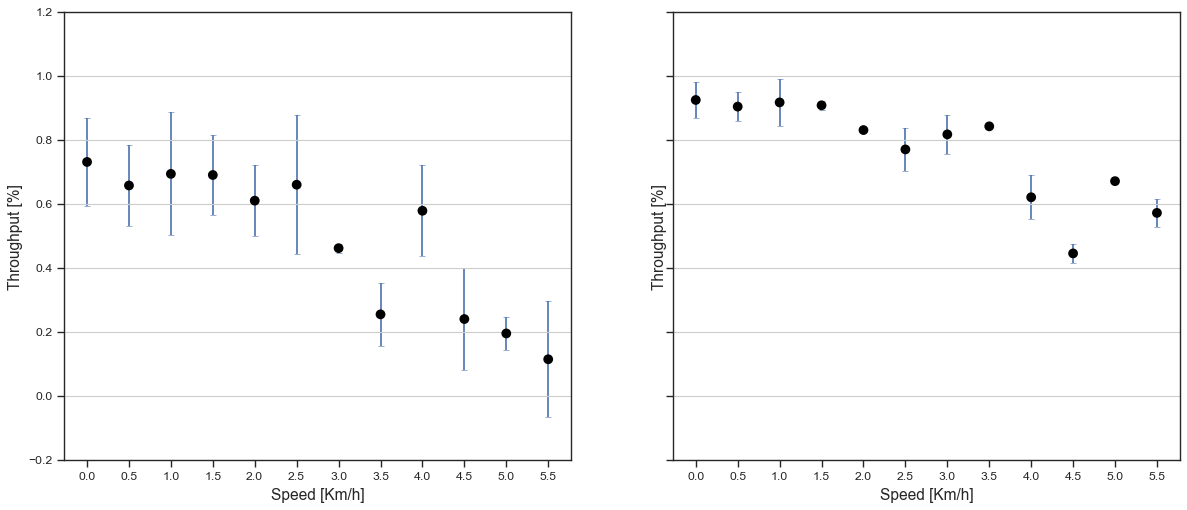}
\caption{Effect of Speed on Throughput for TCP (left) and Disaggregated Memory Modules (right). Error bars show the estimated standard error of the mean (SEM) calculated with at least 5 samples per speed.}
\label{fig:speed}
\end{figure}

Figure \ref{fig:speed} shows experiments comparing data transfers in a Disaggregated Edge (right) versus traditional TCP (WiFi) transfers (left). Data requests in a Disaggregated Egde have a smaller stack and reduced encapsulation results in smaller data sizes and slightly higher throughput\footnote{Note this prototype is not using a real NVM hardware chip, but we simulate its behaviour with an SD card. The simulation means there will be higher latencies and lower throughput than in a setup with real NVM deployed.}. 

Transfer rates for TCP connections present more variability  because the transfers  are  more  susceptible  to connection  transitions between WiFi nodes.

The synchronisation of the different elements worked well at low speed, but it failed for speeds faster than 6Km/h as the protocol did not have the time to converge in the inclusion of the car in the overlay. In some cases, the request is made available to the overlay and the car joins the overlay, but the response cannot be delivered if the car is out of the range of all the memory modules, causing it to crash against obstacles. 

Note that our goal here was not to prove the feasibility of NVM dynamic aerial fabrics for safe self-driving cars, but we rather wanted to show the feasibility of this architecture to reduce the processing needs in reality enhancement scenarios. 

\section{Discussion and Future Work}
\label{sec:discuss}

\subsection{Advantages}


Disaggregated edges can provide differentiating performance for essential infrastructure, such as: reduced VM/container booting times, container acceleration using NVM-supported data replication (instead of building application specific solutions), or live VM migration. This can be done by changing the ownership of remote NVM pages associated with a migrating VM transferred between compute hosts~\cite{Gao16,Lim2012}.

Having a set of resources that can be dynamically put together when needed in very small units of execution means that the distinction between horizontal and vertical scalability disappears, which is comparable to aggregating more resources for their application or service. The illusion of a dynamically expandable virtual memory space where edge nodes route memory requests even when the requester is moving  simplifies data sharing and enables more stability in distributed memory requests.


In the edge, a portion of the resources will likely fail and some execution units will have to be re-executed (unless preemptive techniques are applied) and some of the data in failed memory units will have to be replicated so as to ensure data availability. For instance, the performance of straggler memory nodes (or high latency/unreliable networks) dominates response time, which is very critical in latency-sensitive applications. 


\subsection{Security}

Security is essential in hierarchical or fully disaggregated NVM edge clouds. These are highly dynamic and multi-tenant environments where acceleration of encryption, policy engines and enforcement controls tightly-integrated with networking, and decentralised identity management infrastructures will prove crucial~\cite{Roman2013}. NVM introduces some security considerations of its own. As data is persistently stored in memory, new mechanisms for naming and access control would be needed to prevent the possibility of cross-talk. Also, large scale hardware-accelerated ubiquitous encryption requires local services to help encrypt/decrypt data and new techniques for distributed identity management.

\subsection{Will it Be More Difficult to Develop?}

New programming models are required that will take into account NVM memory locality, coherent access, churn, energy-efficiency, resource constraints, and heterogeneity at the edge of the network.

Thus, system software engineers will have to adapt current operating systems and middleware to hide changes in hardware/firmware of cloud providers~\cite{Shan2018}. Programming abstractions helping with NVM edge memory management. In the case of remote byte-addressable, examples of these mechanisms are: dealing with different coherence domains, durable atomic updates, memory garbage collection/zero-ing, etc.~\cite{Gao16}. 

Consistent updates to remote memory may also be delegated to libraries that free developers from handling functions like memory allocation, leak prevention, type checking, durable transactions and atomic updates. These libraries tend to be general and they can preclude low-level application-specific optimisations and result in conservative ordering constraints~\cite{Nalli2017}.




\subsection{Future Directions}
In order to perform a query or transaction across multiple objects, the application needs to do some extra work. \cite{vaquero2020b} shows how to enable commit based operations in a distributed fabric of NVM memory modules. 

We see a future of edge databases that store each piece of information as a NVM bytearray and databases that operate at the edge. In future works, we will explore how the main building blocks of databases will have to be adapted to work over disaggregated edge clouds.

As shown in our car prototype, disaggregated edges can operate well at human speed, but fall short to support devices on the move at speeds slightly superior to the average human pace. Future research on more efficient aerial protocols and faster overlay convergence is required.

We also plan to benefit from knowledge on the direction and speed of a moving object to predict future location and preemptively allocate data that may be needed in nearby NVMs. This has been explored as a library sitting on top of traditional operating systems such as DAL~\cite{Nemeth2017}. Also, we foresee the exploration of Conflict-free Replicated Data Types (CRDTs) to deal with coherence and concurrent access to shared data \cite{Krol19}. 

\section{Conclusions}
\label{sec:conc}
New hardware technologies like NVM, DSM systems, photonic interconnects, and hardware disaggregation at the edge of the network are reshaping how future services will be supported and built. 

These technologies concur into disaggregated edges, which provide the means for a dynamically expandable virtual memory space were edge nodes route memory requests even when the requester is moving. Disaggregated edges simplify data sharing and enable more stability in distributed memory requests with moving or volatile resources.

We have presented an architectural realisation of the amalgamation of these technologies and proved its feasibility with 2 prototypes: an initial implementation on a NVM hardware module and a simulation of NVM module. 


Some of the main challenges are shared with DSM systems, such as keeping  developers unaware of the complexity, heterogeneity, and high churn rates in unreliable networks with objects moving at high speeds.


\bibliographystyle{ACM-Reference-Format}
\bibliography{sample-base}

\end{document}